\documentclass[showpacs,twocolumn,amsmath,superscriptaddress,prl]{revtex4}
\usepackage{epsfig,verbatim}

\begin{document}

\preprint{EV/18-01-2009}

\title{Correlations in avalanche critical points}
\author{Benedetta Cerruti}
\affiliation{ Departament d'Estructura i Constituents de la Mat\`eria,
  Universitat de Barcelona \\ Mart\'{\i}  i Franqu\`es 1 , Facultat de
  F\'{\i}sica, 08028 Barcelona, Catalonia}
\author{Eduard Vives}
\email{eduard@ecm.ub.es}
\affiliation{ Departament d'Estructura i Constituents de la Mat\`eria,
  Universitat de Barcelona \\ Mart\'{\i}  i Franqu\`es 1 , Facultat de
  F\'{\i}sica, 08028 Barcelona, Catalonia}

\date{\today}

\begin{abstract}
Avalanche dynamics and related power law statistics are ubiquitous in nature,
arising in phenomena like earthquakes, forest fires and solar flares.
Very interestingly, an analogous behavior is associated with many condensed
matter systems, like ferromagnets and martensites. Bearing it in mind, 
we study the prototypical 3D RFIM at $T=0$. We find a finite correlation
between waiting intervals between avalanches and the previous avalanche size. 
This correlation is not found in other models  for avalanches, 
such as  the standard BTW model, but  it is experimentally  
found in earthquakes and in forest fires. Our study suggests that this effect
 occurs  in critical  points which  are  at the  end of  an
 athermal first-order  transition line separating  two behaviors: one
 with high activity from another with low activity.
\end{abstract}



\maketitle

In the  last few  years much experimental  and theoretical  effort has
been devoted  to the study of avalanche  processes. Deep understanding
of the statistical correlations in such stochastic processes is needed
in order  to make advances  towards predictability. The  importance of
the  subject is  beyond discussion  due  to the  many implications  in
natural  disasters   and  social  crises.    Avalanche  processes  are
characterized by extremely fast events whose occurrences are separated
by waiting intervals  without activity.  The magnitudes characterizing
avalanches (energy, size, duration)  are, in most cases, statistically
distributed  according to  a power  law  $p(s) ds  \sim s^{-\tau}  ds$
characterized by  a critical  exponent $\tau$: extremely  large events
hardly  occur, whereas  small events  are  very common.   This is  the
famous Gutenberg-Richter law  for earthquakes. Power-law distributions
have  been found  not  only for  other  large-scale natural  phenomena
ranging   from  solar   flares  \cite{Boffetta1999}   to  forest-fires
\cite{Malamud1998},  but  also  in  laboratories associated  with  many
condensed matter  systems: condensation \cite{Lilly1993}, ferromagnets
\cite{Bertotti1998},    martensitic   transitions   \cite{Vives1994a},
superconductivity \cite{Wu1995}, etc.   However, statistics of waiting
intervals  $\delta$ are  not  often studied,  especially in  condensed
matter.  The  distribution $p(\delta)  d\delta$ has been  described by
different laws including exponentials  and also power-laws.  For solar
flare   statistics  and   earthquake   models,  it   has  been   found
\cite{Sanchez2002,Paczuski2005}   that   the   unavoidable   threshold
definition   (separating   activity   from  inactivity)   alters   the
distribution  $p(\delta)$  and  that  this thresholding  effect  is  a
signature of the existence of correlations \cite{Yang2004,Woodard2004}.
Direct measurement  of the correlations between  waiting intervals and
avalanche  sizes  has  been  obtained  by  measuring  the  constrained
interval    distributions   $p_{prev}(\delta    |    s>   s_o)$    and
$p_{next}(\delta | s'> s_o)$.  These are the probabilities of having a
waiting interval $\delta$,  given that the previous ($s$)  or the next
($s'$) avalanche is larger than $s_0$.  For earthquake and forest-fire
statistics,  while $p_{next}$  has  been found  to  be independent  of
$s_0$, $p_{prev}$ does exhibit  significant changes when varying $s_0$
\cite{Corral2005,Corral2008}.

In this letter we study  some statistical correlations for the 3D-RFIM
at $T=0$ with  metastable dynamics based on  the local relaxation of
single  spins. This  model  was introduced  \cite{Sethna1993} for  the
study of Barkhausen noise in ferromagnets \cite{Bertotti1998} and
acoustic  emission in  martensitic transitions  \cite{Vives1994a}, and
has been  used as  a prototype  for the study  of crackling  noise and
other  avalanche   phenomena  \cite{Sethna2001,Detcheverry2005}.   The
model  is defined  on  a cubic  lattice  with size  $N=L^3$.  At each
lattice site there is a  spin variable $S_i=\pm 1$ $(i=1,\dots, N)$ that
interacts  with  its  nearest   neighbors  (n.n.)   according  to  the
Hamiltonian (magnetic enthalpy):
\begin{equation}
{\cal H} = - \sum_{n.n.} S_i S_j -\sum_{i=1}^N S_i h_i -H \sum_{i=1}^N
S_i.
\end{equation}
The local random fields $h_i$  are Gaussian distributed with zero mean
and standard  deviation $\sigma$. This  parameter not only  allows the
critical  behavior ($\sigma=\sigma_c  \simeq  2.21 \pm  0.01$) to  be
studied   but   also  subcritical   ($\sigma>\sigma_c$),   and
supercritical                ($\sigma<\sigma_c$)               regimes
\cite{Perkovic1999,PerezReche2003}.   This tuning is absent in
so-called  Self-Organized Criticality models like the original
BTW  model  \cite{Bak1987}, in  which  the  critical state  is
reached  after  waiting  for  a  certain time  without  any  parameter
adjustment.   In  the RFIM, the time  variable  is replaced  by  the
external field $H$, which is adiabatically increased from $-\infty$ to
$\infty$.  The system  responds by an increase of  the order parameter
$m \equiv \sum_{i=1}^N S_i /N  $ (magnetization per spin) from $-1$ to
$1$.   The spins  flip according  to the  dynamical rule  $S_i  = sign
(\sum_j S_j + h_i + H)$ (the first sum runs over the n.n. of spin $S_i$), 
which corresponds to a minimization of the local energy.  This
non-equilibrium dynamics leads to  hysteresis and avalanches when many
spins flip  at constant field.  The  avalanche size $s$  is defined as
the number of  spins flipped until a new  stable state is reached.
Avalanches are  separated by field waiting  intervals $\delta$ without
activity.   When $\sigma=\sigma_c$  and $H$  is close  to  $H_c \simeq
1.43\pm0.05$  \cite{Perkovic1999,PerezReche2004b}, the  avalanche size
distribution becomes approximately a  power law $p(s) \sim s^{-\tau}$
(with $\tau \simeq 1.6$ \cite{Perkovic1999}).  The properties of large
(percolating) avalanches  at the  critical point have  been previously
discussed \cite{PerezReche2004b}.  Critical avalanches have been found
to be fractal so that $\langle s \rangle_c \sim L^{d_f}$ with $d_f\sim
2.88$.  When  $\sigma > \sigma_c$,  the avalanche distribution is
exponentially  damped, i.e.   all avalanches  are  negligible compared
to  system size, and the magnetization  $m$ evolves continuously
when the system is infinite.   For $\sigma < \sigma_c$, avalanches are
also  infinitesimally small  compared to  $L^3$, except  for  a unique
infinitely large and compact ($s \sim L^3$) avalanche corresponding to
a  first-order phase transition  between a  phase with  low $m$  and 
a phase with high $m$.  The  first-order transition line can be linearly
approximated  by the equation   
$H_t(\sigma)=H_c\left[1-B'(\sigma-\sigma_c)/\sigma_c\right]$  with
$B'=0.25$. \cite{PerezReche2004b}

\begin{figure}[ht]
\begin{center}
  \epsfig{file=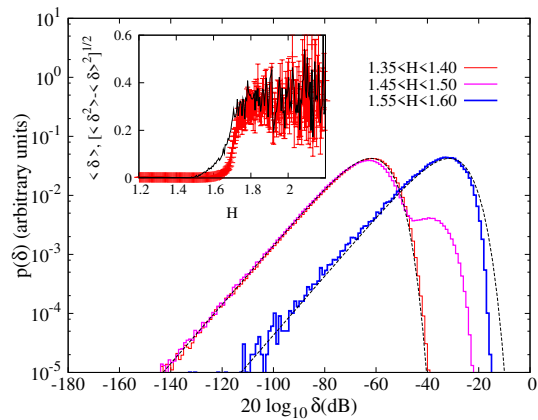,width=8.0cm,clip=}
\end{center}
\caption{\label{FIG1} Log-log plot  of the histograms corresponding to
the   distribution   of  waiting   intervals   $\delta$  for   $L=30$,
$\sigma=2.21$  and  three different  field  ranges,  indicated by  the
legend.   Note  that  the horizontal  scale  is  in  dB and  bins  are
logarithmic.  The  dashed lines show  fits corresponding to  the exact
exponential behavior.  Data have been obtained by averaging over 50000
realizations of  disorder.  The inset  shows the behavior  of $\langle
\delta  \rangle$  (symbols)  and  $\sqrt{  \langle  \delta^2\rangle  -
\langle \delta \rangle^2 }$ (continuous line) as a function of $H$ for
$\sigma=1.95$ and $L=30$.}
\end{figure}

In  this letter  we  will concentrate  on  the analysis  of the  field
waiting intervals  $\delta$, and  their correlation with  the previous
and next  avalanche sizes.  We  have first checked that  intervals are
exponentially distributed.  Fig.~\ref{FIG1} shows the distribution of
waiting  intervals  for  $\sigma=2.21$,  $L=30$  and  three  different
$H$-ranges.   One can  observe that  before and  after  the transition
region  the  distributions are  very  well  described by  exponentials
(continuous lines), whereas in  the transition region the distribution
becomes  a linear mixture  of exponentials.   The mean  value $\langle
\delta  \rangle$,  which is  the  only  parameter characterizing  such
distributions, depends  on $\sigma$ and  $H$, and below  $\sigma_c$ it
exhibits  a  discontinuity  $\Delta  \delta$ when  $H$  increases  and
crosses  the  first-order transition  line  from  the  region of  high
activity  (small  $\langle  \delta  \rangle$)  to the  region  of  low
activity (large $\langle \delta \rangle$  ), as shown in the inset of
Fig.~\ref{FIG1}.   We  can, therefore,  compare  the  behavior of  the
discontinuity  $\Delta \delta$ to  that of  an order  parameter.  Note
that, given the  finite size of the system,  the pseudo-critical point
where  the  mixture  of  exponentials  becomes  a  single  exponential
distribution  will  be located  at  $\sigma>2.21$.   For this  reason,
although data  in Fig.~\ref{FIG1} correspond to  $\sigma_c$, one can
still observe a range of fields with the two-peak distribution.

Besides the  histogram analysis, we have numerically  checked that for
all  $\sigma$ and  $H$, $\langle  \delta \rangle  \sim  \sqrt{ \langle
\delta^2\rangle - \langle  \delta \rangle^2 }$,\footnote{Except on the
transition  line where  we  expect $\sqrt{  \langle \delta^2\rangle  -
\langle \delta \rangle^2 } \sim \sqrt{ \langle \delta \rangle ^2 - k_0
\Delta \delta ^2}$  with $k_0=$ constant.}  which is  more evidence of
the exponential  character of $p(\delta)$. The continuous  line in the
inset  of  Fig.~\ref{FIG1}  shows  this  agreement,  except  for  some
deviations close to the transition line due to finite-size effects.
%

\begin{figure}[ht]
\begin{center}
  \epsfig{file=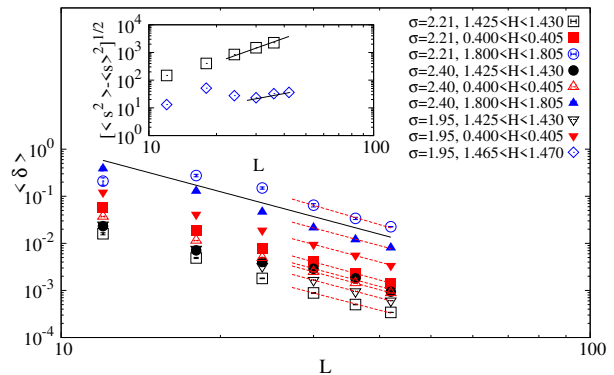,width=8cm,clip=}
\end{center}
\caption{\label{FIG2}  Log-log plot  of the  average  waiting interval
$\langle  \delta \rangle$ as  a function  of the  system size  $L$ for
different values of  $\sigma$ and $H$ as indicated  by the legend. The
continuous  line shows  the  behavior $L^{-3}$  and the  discontinuous
lines are fits  to the last 2-3 points for each  series of data.  Data
are  averaged over 50000  disorder realizations.  The inset  shows the
behavior of  $\sqrt{\langle s^2 \rangle-\langle s  \rangle^2}$ for the
values of $\sigma$  and $H$ indicated.  The continuous  lines show the
behavior $L^{d_f}$ (above) and $L^{3/2}$ (below).}
\end{figure}

For  the  following  discussions  it  is interesting  to  analyze  the
finite-size dependence of $\langle  \delta \rangle$. A simple argument
can  be used  to state  that far  from the  critical point,  since the
correlation  length is  finite, the  probability for  an  avalanche to
start  when the  field is  increased by  $dH$ is  proportional  to the
number  of triggering  sites and  thus  to $L^3$.   This implies  that
$\langle \delta \rangle  \sim L^{-3}$.  Fig.~\ref{FIG2} shows examples
of  this  behavior for  different  values  of  $\sigma$ and  $H$.   At
criticality,  the  correlation  length  diverges  and  the  triggering
argument may be too naive.  Nevertheless, as shown in Fig.~\ref{FIG2},
numerical simulations indicate that  the exponent is always very close
to  $3$. Uncertainties in  $H_c$ and  $\sigma_c$ do  not allow  for an
accurate  enough  finite-size  scaling  analysis  to  determine  small
variations.  For the discussions here the exact value of this exponent
is not  needed and  we will assume  that $\langle \delta  \rangle \sim
L^{-z}$ with $z \sim 3$.
%

\begin{figure}[ht]
\begin{center}
  \epsfig{file=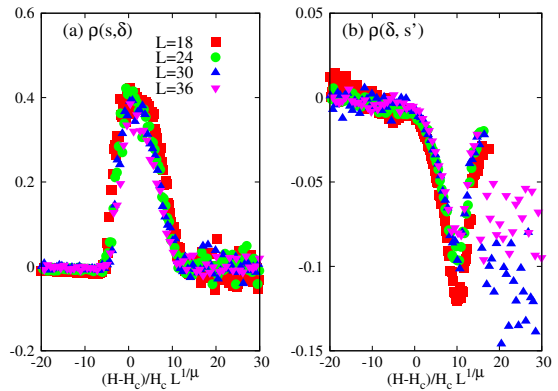,width=7.5cm,clip=}
\end{center}
\caption{\label{FIG3}    Correlation   functions   defined    by   Eq.
(\ref{co3}) and  (\ref{co4}) as functions of the  scaling variable for
$\sigma=2.21$ and  different system sizes as indicated  by the legend.
Data correspond to averages over 50000 disorder realizations and field
intervals with  size $\Delta H=  0.005$.  Note the  different vertical
scale in the two figures.}
\end{figure}

Let us now focus on the  study of correlations and consider a sequence
of two  avalanches, the first with  size $s$, then  a waiting interval
$\delta$,  and the  next  avalanche  with size  $s'$.   We define  the
following two correlation functions:
\begin{eqnarray}
 \label{co3}
\rho_{s,  \delta} & =  & \frac{\langle  s \delta  \rangle -  \langle s
\rangle \langle \delta \rangle} {\sqrt{\langle s^2 \rangle - \langle s
\rangle ^2}  \sqrt{\langle \delta^2  \rangle - \langle  \delta \rangle
^2}}\\
 \label{co4}
\rho_{\delta, s'} &  = & \frac{\langle s' \delta  \rangle - \langle s'
\rangle \langle \delta \rangle}  {\sqrt{\langle s'^2 \rangle - \langle
s' \rangle ^2} \sqrt{\langle \delta^2 \rangle - \langle \delta \rangle
^2}}
\end{eqnarray}
Fig.~\ref{FIG3}  shows examples  of  the behavior  of the  correlation
functions for different system sizes and $\sigma=\sigma_c \simeq 2.21$
as a function  of the scaling variable $v=\left  [(H-H_c)/H_c \right ]
L^{1/\mu}$  that measures  the distance  to the  critical  point.  The
exponent    $\mu=1.5$   has   been    taken   from    the   literature
\cite{PerezReche2004b}.  (Note  that this scaling  variable only takes
three values in the thermodynamic limit $v=\pm \infty, 0$).

For the understanding of the behavior of the correlation functions for
increasing  system size,  one  must first  discuss  what the  expected
behavior     of    a     generic     correlation    function     $\rho
(H,\sigma=\sigma_c,L)$ close to  $H_c$ is likely to be.   It should be
borne in  mind that correlations  are, by definition,  bounded between
$-1$  and  $1$.  Therefore  no  critical  divergences  can occur  with
increasing $L$.  Consequently, any critical correlation either goes to
zero or  tends to  a constant  value. In this  second case,  it should
exhibit     scaling     behavior     $\rho     (H,\sigma_c,L)     \sim
\hat{\rho}(\sigma_c, v)$.

The first observation from Fig.~\ref{FIG3} is that $\rho_{\delta, s'}$
is much smaller than $\rho_{s, \delta}$ not only at the critical point
($v=0$), but also for other values of the field. In addition, the data
shows  that $\rho_{\delta, s'}$  systematically decreases  in absolute
value with  increasing system size.   Therefore the numerical  data is
consistent with  a vanishing $\rho_{\delta, s'}$  in the thermodynamic
limit.

The  second  important  observation  is that  correlation  between  an
avalanche size and the  next waiting time, $\rho_{s, \delta}$ exhibits
a constant value $\sim 0.4$ at  the critical point. The overlap of the
curves is  rather good, specially if  one takes into  account the fact
that  even at  the critical  point there  are  non-critical avalanches
which  may slightly perturb  the scaling  function behavior.  The fact
that  the scaling  function in  Fig.~\ref{FIG3}(a)  goes to  zero for
$v\rightarrow \pm  \infty$ indicates that the  finite correlation only
survives exactly at  the critical point, but vanishes  for fields both
above and  below.  We would like to  note that the result  of a finite
correlation  $\rho_{s,  \delta}$  is  not in  contradiction  with  the
Poissonian character for the triggering instants of the avalanches.

\begin{figure}[ht]
\begin{center}
  \epsfig{file=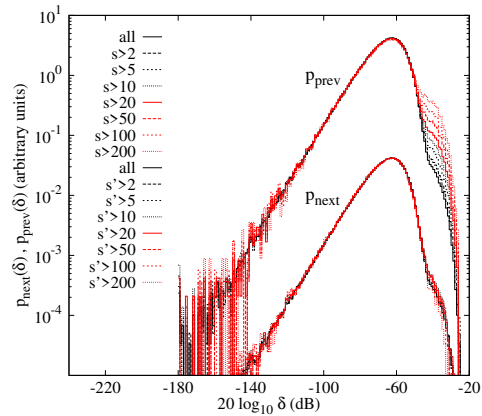,width=8cm,clip=}
\end{center}
\caption{\label{FIG4} Log-log  plot of the  restricted distribution of
intervals $\delta$ given that  the previous (next) avalanche is larger
than $s_0$. $p_{prev}$  has been displaced two decades  up in order to
clarify the picture.}
\end{figure}

A second way to go deeper  into the understanding of the nature of the
correlations is the direct  measurement of the restricted distribution
of  intervals $p_{prev}(\delta  | s>s_o)$  and $p_{next}(\delta  | s'>
s_o)$.  Fig.~\ref{FIG4} shows, as an example, the dependence of these
two distributions for $L=30$, $\sigma=2.21$ and $1.40<H<1.45$.
%
%
The  distribution  of  intervals  $p_{prev}(\delta |  s>s_o)$  clearly
exhibits  a  dependence  on   the  previous  avalanche  size,  whereas
$p_{next}(\delta | s'> s_o)$ is independent of $s_0$. Thus, the larger
the  size  of  an  avalanche,  the larger  the  probability  that  the
following waiting  interval is  large.  We should  note at  this point
that  a  similar  same  causal  dependence  has  been  found  for  the
statistics       of      earthquakes       and       forest      fires
\cite{Corral2005,Corral2008},  but with a  different sign.  The larger
the size  of an earthquake, the  smaller the waiting time  to the next
event.

The  origin of  such correlations  in our  RFIM can  be  understood by
noting that, for a finite  system below $\sigma_c$, after an avalanche
with a  large size (of  the order $L^3$),  one can ``guess''  that the
system has jumped  to the low-activity region and,  therefore, one can
``predict'' that  the next  interval $\delta$ will  be large.   We can
provide  an heuristic  argument of  why such  finite-size correlations
vanish in  the thermodynamic limit, everywhere except  at the critical
point.   Let us  analyze the  behavior of  the numerator  and  the two
square roots  in the denominator  in Eq.  \ref{co3}. First  note that,
given the fact that $p(\delta)$  is exponential, the fluctuations of $
\sqrt{\langle \delta^2 \rangle - \langle \delta \rangle ^2}$ behave as
$\delta  \sim L^{-z}$.   The  fluctuations of  $s$ behave  differently
below $\sigma_c$ and at $\sigma_c$. If we consider an interval $\Delta
H$ that crosses the first-order transition region, below $\sigma_c$ we
will have a number of avalanches proportional to $L^3$ contributing to
the averages. Among  these, most will display a  small size $\sim L^0$
but  one will  have  a size  $\sim  L^3$.  When  computing $\langle  s
\rangle$  we get a  $\langle s  \rangle \sim  L^0$ behavior,  but when
computing $\sqrt{\langle s^2 \rangle -  \langle s \rangle ^2}$ we will
get  $L^{3/2}$,  as  shown  in  the  inset  of  Fig.~\ref{FIG2}  for
$\sigma=1.95$ and the interval $1.465<H<1.470$ crossing the transition
line.  Exactly at criticality,  the distribution of avalanches becomes
a  power-law. This  means that  the averages  $\langle s  \rangle$ and
$\langle  s^2   \rangle$  should   be  computed  by   integrating  the
distribution  from $1$  to  the  largest avalanche  size  which has  a
fractal dimension  $s_{max} \sim L^{d_f}$  \cite{PerezReche2004b}.  By
integration, one  trivially gets $\langle s \rangle  \sim L^{d_f}$ and
$\langle s^2 \rangle \sim L^{2 d_f}$.  The fluctuations therefore will
go  as $  \sqrt{\langle s^2  \rangle -  \langle s  \rangle  ^2} \simeq
L^{d_f}$, as shown in the inset of Fig.~\ref{FIG2}.

To study the  numerator in Eq.  (\ref{co3}) one should note
that $  \langle s  \delta \rangle -  \langle s \rangle  \langle \delta
\rangle  = \langle  s' \delta'  \rangle -  \langle s'  \rangle \langle
\delta \rangle  = \langle s' (\delta'-\delta)  \rangle$.  This average
measures   the  avalanches   that  carry   an  associated   change  in
$\delta$. For $\sigma<\sigma_c$, among  the set of $L^3$ avalanches in
a  interval  $\Delta  H$,  there  is one  avalanche  with  size  $L^3$
associated with a change $\Delta \delta \sim L^{-z}$.  The rest of the
avalanches   have   a   size    $L^0$   and   carry   no   change   in
$\delta$. Therefore  the numerator in Eq.  \ref{co3}  goes as $L^{-z}$
and  consequently,  below  $\sigma_c$,  $\rho(s,\delta)  \sim  L^{-z}/
(L^{3/2} L^{-z}) \sim L^{-3/2} \rightarrow 0$.

For $\sigma = \sigma_c$ the  behavior of the numerator in (\ref{co3})
is much  more intricate since  different kinds of  critical avalanches
exist  close  to   the  critical  point.   Apart  from   a  number  of
non-critical avalanches, there is an infinite number ($\sim L^\theta$)
of   spanning   avalanches   and   another   infinite   number   $\sim
L^{\theta_{nsc}}$ of critical non-spanning avalanches. It is difficult
to argue which of these avalanches have an associated $\Delta \delta$.
We  cannot provide  a definite  argument, but,  at least,  the product
$\langle s \rangle \langle \delta \rangle$ (appearing in the numerator
of  Eq.\ref{co3})  behaves  as  $L^{d_f}  L^{-z}$.   Therefore  it  is
plausible that at $\sigma_c$, $\rho(s,\delta) \sim L^{d_f-z}/ (L^{d_f}
L^{-z}) \sim 1$,  which justifies the finite value  of the correlation
$\rho(s,\delta)$ found numerically, as shown in Fig.~\ref{FIG3}.

In order  to compare with the  avalanche models based on  SOC, we have
performed  an   analysis  of   the  same  probability   densities  and
correlations for the 2D-BTW model. In this case, the waiting times are
discrete since  they are  identified with the  number of  grains added
before  a  new avalanche  starts.   For  large  enough systems,  these
waiting   intervals  are  distributed   according  to   the  geometric
distribution  which  is  the   discrete  version  of  the  exponential
distribution  and all  the two  correlation functions  (\ref{co3}) and
(\ref{co4}) clearly vanish.

In summary, we have numerically studied the $T=0$ RFIM with metastable
dynamics as  a prototype for  avalanche processes in  condensed matter
systems which display an  underlying first-order phase transition. The
sequence  of avalanches  and  waiting  times can  be  considered as  a
compound  Poisson   process  since   waiting  intervals  tend   to  be
exponentially  distributed.   Nevertheless,  correlations between  the
avalanche size and  the next waiting time exist  at the critical point
in the thermodynamic limit.  Such a causal correlation $\rho(s,\delta)
\neq  0 $  has been  found  experimentally in  earthquakes and  forest
fires, although  with a different sign. This  difference could easilly
be  explained by considering  a distonintuity  $\Delta \delta  <0$. An
experimental challenge for the future  is to look for these effects in
laboratory   experiments  on   condensed  matter   systems  exhibiting
avalanches (Barkhausen noise, acoustic emission in martensites, etc).

This work  has received financial support from  CICyT (Spain), project
MAT2007-61200,     CIRIT     (Catalonia),    project     2005SGR00969,
B.C. acknowledges  the hospitality  of ECM Department  (Universitat de
Barcelona)  and a  grant  from  Fondazione A.  Della  Riccia. We  also
acknowledge fruitful discussions with A.Corral, J. Vives and A. Planes.

\end{document}